\documentclass[aps,prl,groupedaddress]{revtex4}%
 \pdfoutput=1
\usepackage{graphicx}

\newcommand{\bild}[5]{\begin{figure}[#3]
                        \centering
                        \includegraphics[width=#2]{#1}
                        \sl\caption{#4}
                        \label{#5}
                        \end{figure}}

\usepackage{xspace}

\newcommand{\Figref}[1]{Fig.~\ref{#1}}
\newcommand{\eqnref}[1]{Eq.~\ref{#1}}

\newcommand{\enquote}[1]{``#1''}

\begin{document}
\pagestyle{plain}

\title{Ultra-fast transistor-based detectors for precise timing of near infrared and THz signals}

\author{S. Preu,$^{1,*}$ M. Mittendorff,$^{2,3}$  S. Winnerl,$^2$ H. Lu,$^{4}$ A. C. Gossard$^{4}$, and H. B. Weber$^1$}
\address{$^1$Chair for Applied Physics, Univ. of Erlangen-Nuremberg, Germany, *email: sascha.preu@physik.uni-erlangen.de}
\address{$^2$Technical University Dresden, Dresden, Germany}
\address{$^3$Helmholtz-Zentrum Dresden-Rossendorf, Germany}
\address{$^4$Materials Department, University of California, Santa Barbara, California, USA}

\begin{abstract}
A whole class of two-color experiments involves intense, short Terahertz radiation pulses. A fast detector that is sensitive and able to resolve both near-infrared and Terahertz pulses at the same time is highly desirable. Here we present the first detector of this kind. The detector element is a GaAs-based field effect transistor operated at room temperature. THz detection is successfully demonstrated at frequencies up to 4.9 THz. The THz detection time constant is shorter than 30 ps, the optical time constant is 150 ps. This detector is ideally suited for precise, simultaneous resolution of optical and THz pulses and for pulse characterization of high-power THz pulses up to tens of kW peak power levels. The dynamic range of the detector was as large as 65 $\pm$ 3 dB/$\sqrt{Hz}$, enabling applications in a large variety of experiments and setups, also including table-top systems.
\end{abstract}
\maketitle

\section{Introduction}
Progress in Terahertz (THz) research (100 GHz-10 THz) has yet reached a technological level that is attractive for a huge variety of experiments and real-life applications. The established fields of applications include security and imaging,\cite{Pfeiffer2010,CW-ImagingRoskos,FET1THz} spectroscopy,\cite{ZhangReview02,Tonouchi07} optical pump-THz probe and, since very recently, THz pump-optical probe experiments\cite{TaniTableTopTHzPumpOptProbe}. The latter became attractive tools for studying ultrafast processes in matter. Pump-probe experiments include probing the electronic structure,\cite{HuberExcitons} carrier relaxation experiments,\cite{HuberParticleInteractions, GeorgeGrapheneCarrRelax} high harmonic generation, and perturbative nonlinear phenomena.\cite{BenHighHarmonic,WagnerResonantSidebandFEL} Due to the high speed of light, a small length delay between the two pulses allows for probing physics on a sub-ps timescale. In many cases, modified THz time domain systems are used\cite{HuberExcitons,GeorgeGrapheneCarrRelax} where the same optical beam is used for generating both the THz pulse and the optical probe pulse. 
Many experiments require extremely high THz fields beyond the capabilities of table top systems. Such experiments include studies of ponderomotive forces \cite{Ponderomotive}, non-linear effects generated by intense THz radiation, high order sideband generation \cite{BenHighHarmonic,WagnerResonantSidebandFEL} or the Autler Townes effect \cite{SherwinAutlerTownes,WagnerAutlerTownes}. The requirement of strong fields is based on the nature of these experiments: The THz electric field must be comparable to the (built-in) static electric fields of a sample under test, and thus modifying the physics beyond a simple perturbation. For such experiments, high power THz facilities are used, such as free electron lasers (FEL). An optical pulse (in most cases a near-infrared (NIR) laser pulse) has to be synchronized to the FEL pulse. Since the NIR laser and the FEL are independent entities, this requires not only precise locking techniques, but in particular a measurement technique that allows for timing of both pulses at a picosecond time scale. A fast detector is required that is both sensitive to the optical and the THz FEL pulse. There exists a large variety of THz detectors, such as Golay cells, pyroelectric detectors\cite{ZhangReview02,PyroelCamera,FELpyro} or photo-accoustic power meters (Thomas Keating power meter), fast Schottky diodes \cite{FrequenzSchottkyCW}, the superlattice detector, \cite{WinnerlSuperlatt} or photon drag detectors \cite{photonDragDet}. However, none of these is able to detect and resolve a high power THz pulse and an optical pulse synchronously. Furthermore, the detectors have to be able to tolerate the extraordinarily high peak power levels well beyond the kW range and have to be linear at the same time. Such a detector is lacking so far.

We developed a detector that fulfills the aforementioned requirements of ultra high speed, sensitivity to both optical and THz pulses, and high damage threshold. It is based on an antenna-less, large area field effect transistor (LA-FET) used as rectifier. Recently, we demonstrated that such detectors can be used for pulse shape characterization and for power measurements at FELs \cite{MeinFETFEL} at 0.24 THz. We could specify an upper limit for the detector time constant of 10 ns. 

In this paper, we present an improved device and measurements showing a detector time constant below 30 ps as well as synchronous detection of an NIR signal with a similar time constant. The results were obtained at the free electron laser FELBE (Helmholtz-Zentrum Dresden-Rossendorf) with a FEL pulse duration below 33 ps and read out electronics permitting measurements up to a bandwidth of 30 GHz. 
Remarkably, a single detector covers an extremely large frequency span from the low end of the THz band up to at least 5 THz. Although the detection scheme is designed for high power applications, the noise floor should be low enough to permit applications in (lower power) table top systems. Further applications include THz pump-THz probe experiments.

\section{Detector layout and setup} 
The rectifying effect in field effect transistors (FETs) is well known as an efficient method for detection of THz radiation \cite{Pfeiffer2010,ShurAlt,MontpellierII_Mixer,GregIchPRL,GregTransmLineMod}. It is based on a non-linearity of the current within the gated part of the channel. Simply speaking, a THz signal penetrating into the channel will modulate both the carrier velocity, $v(t)\sim U_{THz}(t)$, and the carrier concentration, $n^{(2D)}(t)\sim U_{THz}(t)$, at the same time\cite{MeinFETTheo,ShurRev,AlvydasSensors}. This results in a source-drain current $j_{SD}(t)\sim v(t)n^{(2D)}(t)\sim (U_{THz}(t))^2$. This features a rectified DC component that is proportional to the incident THz power, since $P_{THz}=1/(2R_A)(U_{THz}(t))^2$, where $R_A$ is the radiation resistance of the receiving area or antenna. 
It should be noticed that the transistor is not capable to amplify the THz signal. The rectification effect, however, remains highly efficient far above $f_\tau$ and $f_{max}$, the characteristic cut-off frequencies for current and power amplification \cite{AlvydasSensors}. 
A detailed derivation and description of the detection principle can be found elsewhere \cite{MeinFETTheo,ShurRev,AlvydasSensors}.

Typically, (small) FET rectifiers are attached to an antenna in order to improve the coupling efficiency of the THz power to the detector. The THz power is then concentrated on the few micrometer-scale FET. For the high power levels achieved with FELs, however, this would lead to saturation of the detector or even its destruction. 
Therefore, we developed the large area field effect transistor detector (LA-FET). The power is distributed over a large number of FET mesas arrayed on an area comparable to or even larger than the THz wavelength, resulting in a high damage threshold.\cite{MeinFETFEL} The sample layouts investigated in this study are illustrated in \Figref{Fig1}. The fundamental unit is a long but narrow mesa. Source, gate and drain electrodes are a very wide, but the FET channel is short (aspect ratio 1 mm:19$\mu$m). The $N$ unit cells (c.f. \Figref{Fig1} c) are connected in parallel.

\bild{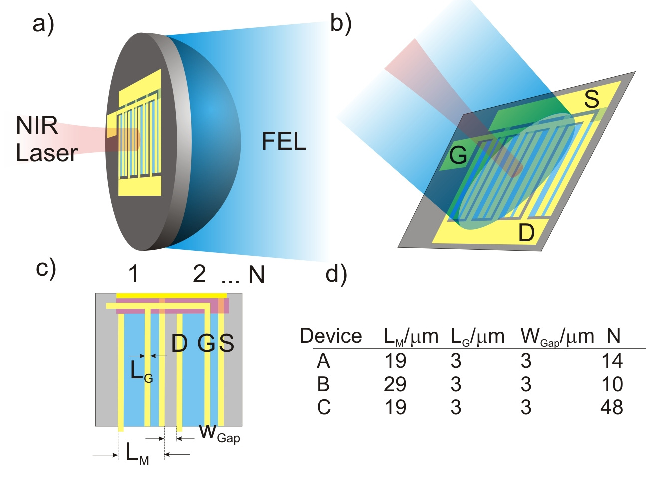}{11 cm}{h}{Sample layout of the FET detectors. a) Silicon-lens coupled devices (A,B). The THz power is coupled through the silicon lens. The optical beam is incident from the other side. b) Free space coupled device (C). Optical and THz beam are both incident for the air-side on the sample. c) Schematic top view. The wiring of the source and gate electrodes is electrically insulated using a SiO$_2$ spacer layer (indicated by the red semi-transparent layer). d) Key dimensions of the sample layouts. }{Fig1}

In this geometry, both the access resistance,$R_{acc}$, and the resistance of the rectifying part of the channel are extremely low. This results in a very short RC time constant. Consequently, the extrinsic speed is extremely high. Since the penetration depth of the THz wave into the gated region (i.e. the effective rectification length) is in the few 100 nm range \cite{MeinFETTheo}, the intrinsic transport time is also extremely short.

The devices were designed to match a diffraction-limited THz spot with a diameter of $2\rho=1.22 \lambda_{THz}/(NA)$ for a numerical aperture $NA=n\times0.25$ for the longest wavelength of 240 $\mu$m (1.25 THz) at FELBE. The index of refraction is $n=1$ for the free space device (C) and $n=n_{Si}=3.4$ for the silicon lens coupled devices (A and B). The active area is (345 $\mu$m)$^2$ for devices A and B and (1.17 mm)$^2$ for device C. Due to processing limitations, the size of device C was chosen slightly smaller with a size of (1 mm)$^2$. The samples and the setup allow for co-parallel optical and THz pulses and for anti-parallel optical and THz pulses as illustrated in \Figref{Fig1} a) and b). A 30 GHz oscilloscope (Tectronix DSA8200) is connected to the source-drain (SD) port of the LA-FET. A bias tee allows for DC biasing the source drain by $U_{SD}$. The gate can be pinched off by an additional DC source-gate (SG) bias, $U_{SG}$. The repetition rate of the FEL is $\nu_{rep}$=13 MHz, while the repetition rate of the optical Ti-sapphire system is 78 MHz. Optical and THz pulses are phase locked resulting in an FEL pulse at every sixth optical pulse. The phase can be freely chosen by an electronic delay. The FEL pulse duration is $\tau_{pls}=$33 ps for the longest investigated wavelength of 230 $\mu$m (1.3 THz) and becomes shorter at shorter wavelengths. The pulse duration of the NIR pulse around 780 nm is 1 ps.

\bild{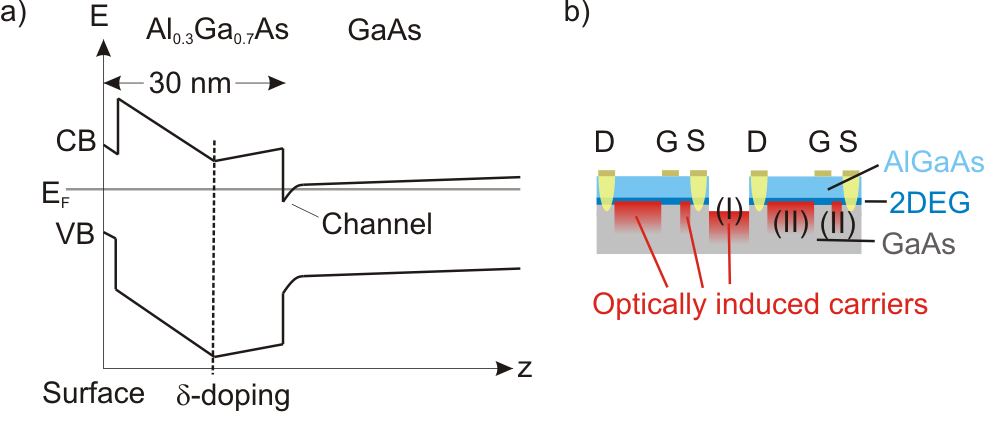}{10 cm}{h}{a) Sample structure of the (Al)GaAs HEMT. The carrier mobility at room temperature is 6700 cm$^2$/(Vs), the carrier concentration is 7$\times$ 10$^{11}$ /cm$^2$. b) Side view of the mesas. Optically excited regions are indicated in red. In area (I), the substrate between two mesas absorbs the NIR laser pulse. Area (II) specifies the absorbing part of the mesa. }{Fig2}

The depletion mode n-FET sample structure is illustrated in \Figref{Fig2} a). The remotely doped high electron mobility transistor (HEMT) consists of a 30 nm Al$_{0.3}$Ga$_{0.7}$As barrier and a GaAs channel. Due to remote doping, the carrier mobility in the channel remains very high, reaching values of 6700 cm$^2$/(Vs) at room temperature, where all measurements are carried out. The samples were processed with electron beam lithography in order to allow for rapid prototyping. 


\section{Results and discussion}

\bild{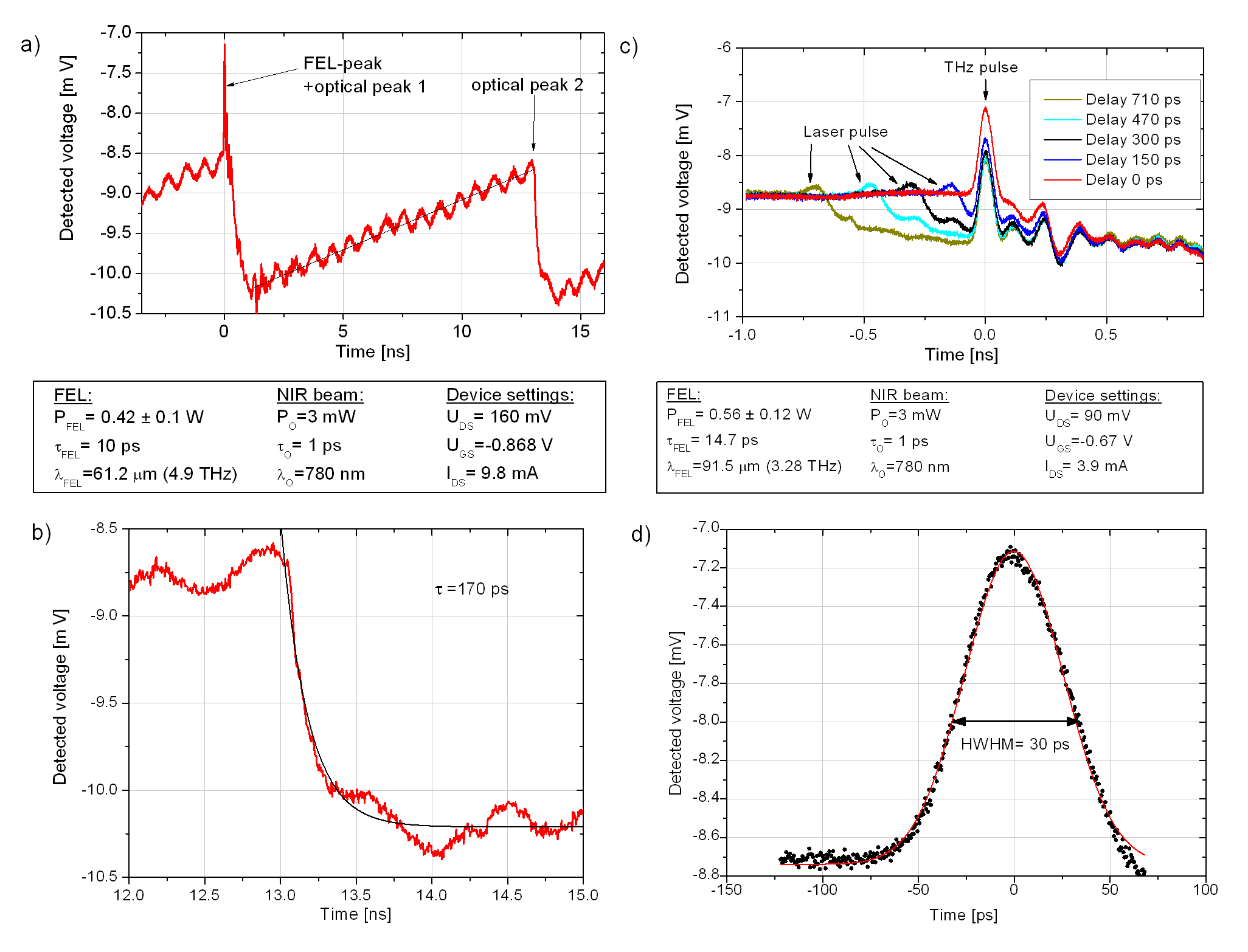}{15 cm}{h}{ a) Sample A, operated at 4.9 THz ($\lambda_{THz}=$61.2 $\mu$m)
: the first optical pulse is synchronized with FEL pulse. For the subsequent NIR pulse, there is no THz FEL pulse since there is only one FEL pulse ($\nu_{rep}=13$ MHz) every 6 optical pulses ($\nu_{rep,o}=78$ MHz). The experimental parameters are summarized in the figure. $P_{FEL}$ is the average power recorded with a thermal power meter. The threshold bias is $U_{thr}=$-1.1 V. The black line indicates the current relaxation between optical pulses. b) Exponential fit of the pulse fall time of the second optical pulse in a), resulting in a 1/e fall time of  $\tau=$170 ps. c) Several measurements with sample B, operated at 3.28 THz ($\lambda_{THz}=$91.5 $\mu$m)
: The optical pulse (laser pulse) has been delayed electronically in several steps with respect to the THz pulse. The delays were extracted from the measurement and are in agreement with the performed electronic delay.  d) Gaussian fit of the measured FEL pulse for zero delay (red graph at t=0 ns in subfigure c).}{Fig3}

\Figref{Fig3} illustrates detection of both optical and THz pulses at the same time for samples A and B. The experimental parameters are listed in the figure caption. The 30 GHz oscilloscope was connected to the drain source contact of the FET. In order to reduce dark currents, the gate bias (relative to the threshold bias) $U_G$ was reduced, increasing the mesa resistance. The optical 1/e fall time at the onset of the optical signal has been determined to $\tau$=170 $\pm$ 20 ps (3 dB fall time $\tau_{3dB}$=120 ps) for sample A at 4.9 THz at a SD bias of $U_{DS}$=160 mV and $\tau$=150 $\pm$ 10 ps (3 dB fall time $\tau_{3dB}$=100 ps) for sample B at 90 mV bias. The relaxation time between optical pulses, indicated by the solid line in \Figref{Fig3} a), is much longer.


We first discuss the response to the NIR laser pulse. 
There are two mechanisms that lead to the optical response of the FETs when the device is subject to a small source-drain bias: (I) The substrate in the etched trenches between neighbouring mesas consists of unintentionally doped, highly resistive GaAs. An optical pulse will generate free carriers that \textit{lower the resistance between neighboring mesas} (see \Figref{Fig2} b). (II) additional carriers are generated in the channel, \textit{altering the threshold bias}.
The measured optical fall times are limited by the transport time of the photogenerated carriers. For a larger bias, photo-generated carriers are more efficiently separated, resulting in a shorter time constant. With the carrier mobility of $\mu\sim$ 6700 cm$^2$/(Vs) (undoped GaAs or in the FET channel), the transport length can be estimated from the Drude model by

\begin{equation}
v=s/\tau_{3dB}=\mu U_{SD}/s \Rightarrow s=\sqrt{\mu U_{DS}\tau_{3dB}}.
\end{equation}

This results in a transport length of $s=$ 3.6 $\mu$m for sample A and 2.5 $\mu$m for sample B. These values are very close to the gap size between the mesas and the gate length of $W_{Gap}=L_G$=3 $\mu$m (c.f. table in \Figref{Fig1}) but much shorter than the mesa length, $L_M$. Obviously both mechanisms contribute to the optical response. By analysis of \Figref{Fig3} a), we conclude that the dominant mechansim, however, is the reduction of the threshold bias (i.e. mechanism II): For absorption between the mesas (mechanism I), the onset of the response should be instantaneous, since the carriers are generated close to the contact pads and the optical pulse time is much shorter than the resolution limit of the oscilloscope. The gap should return to high resistance within about 100 ps-200 ps, i.e. the required time to remove both optically generated electrons and holes. This does not fit to the long relaxation time after the optical pulse shown in \Figref{Fig3} a), black line. Therefore, mechanism (II) is identified to be dominant for the optical response: electron-hole pairs are generated in the mesa structure, except in the gated area which is shadowed by the gate metal. Since the devices were operated close to the threshold bias, most of the applied source drain bias drops at the gated region. Carriers generated close to the gate are drifting into the gated region, resulting in a lowered gate resistance. The pulse fall time (i.e. the fast response to optical excitation) is the time required by electrons to populate the gate. This is in the 100 ps range, in agreement with the measured values. Remaining carriers in the ungated part of the mesa will slowly recombine on the ns scale or be withdrawn to the contacts, resulting in a comparatively long relaxation time after the pulse. The long relaxation time is visible in the measurement in \Figref{Fig3} a) by the non-zero slope of the detection signal, indicated by the black line in \Figref{Fig3} a). 

In both \Figref{Fig3} a) and c), a smaller echo of the FEL pulse with opposite sign is visible in the data (c.f. \Figref{Fig3} c) at 0.3 ns). This echo is due to an RF reflection in the wiring of the LA-FET to the coaxial connection to the oscilloscope. This feature varies from sample to sample. We further see some ringing in \Figref{Fig3} a) and a small positive peak immediately after the optical excitation in \Figref{Fig3} c). These features indicate non-linear effects and may be related to plasmonics. Several papers report on THz generation by optical excitation of FETs in sub-micrometer length channels \cite{OptBeatingDetection, APLOptGatedHEMT}. In order to suppress excessive generation of plasmons which may generate non-linear interaction between the FEL pulse and the near infrared (NIR) pulse, we have chosen long gates of 3 $\mu$m and an NIR pulse duration as large as 1 ps, generating frequency components mainly below 0.5 THz. Since plasmons are only efficiently generated above 0.5 THz in short channels, we do not expect significant generation of plasmons by the optical beam. Therefore, the measured shape, width, and amplitude of the FEL-pulse remain almost unaffected by the optical pulse. The small positive peak in \Figref{Fig3} c) and the ringing in \Figref{Fig3} a), however, requires further experimental and theoretical investigations. It may be the remainder of a plasmonic response.

The detected FEL pulse HWHM in \Figref{Fig3} a) are 20 $\pm$ 0.5 ps (pulse duration 10 ps) for the measurement for sample A and 30 $\pm$ 0.5 ps (pulse duration 14.5 ps) for sample B (\Figref{Fig3} c) and d)). Since the oscilloscope rise time is 17 ps, these values already represent a convolution of the FEL pulse duration, the oscilloscope rise time and the detector time constant. The measured values therefore only represent upper limits for the detector time constant. 

There are many options for improving the performance of the LA-FET. An obvious parameter is the gate bias. In theory, the responsivity should be highest close to pinch-off, $U_G\rightarrow 0$V. In real devices, however, both thermal noise and detection time constant increase for small gate biases, $U_G$. 
 The responsivity of device C increased at lower gate biases, peaking at $U_G=0.25$ V in agreement with impedance matching of the device to the 50 $\Omega$ load. In contrast, the smaller devices A and B showed a decrease in responsivity with lower gate biases (i.e. closer to the threshold bias). Since lower gate biases increase the device impedance, the performance of the smaller devices was already limited by access resistances, resulting in a frequency-dependent and gate bias dependent reduction of the responsivity\cite{MeinFETTheo}. Exact DC values for the access resistance could not be directly measured since they were below the measurement limit of 10 $\Omega$. Using the carrier concentration and mobility obtained from Hall measurements and contact resistance measurements from smaller devices, we estimate a total DC access resistance of 6 $\pm$ 2 $\Omega$ for sample A, of 10 $\pm$ 2 $\Omega$ for sample B, and only  0.6 $\pm$ 0.2 $\Omega$ for sample C. The AC resistance is even significantly lower since the contact resistance is capacitively shorted. The extremely small values for the access resistance are of crucial importance for fast measurement at tens of GHz sampling rates since it reduces any RC time constants in the detection circuit.

\bild{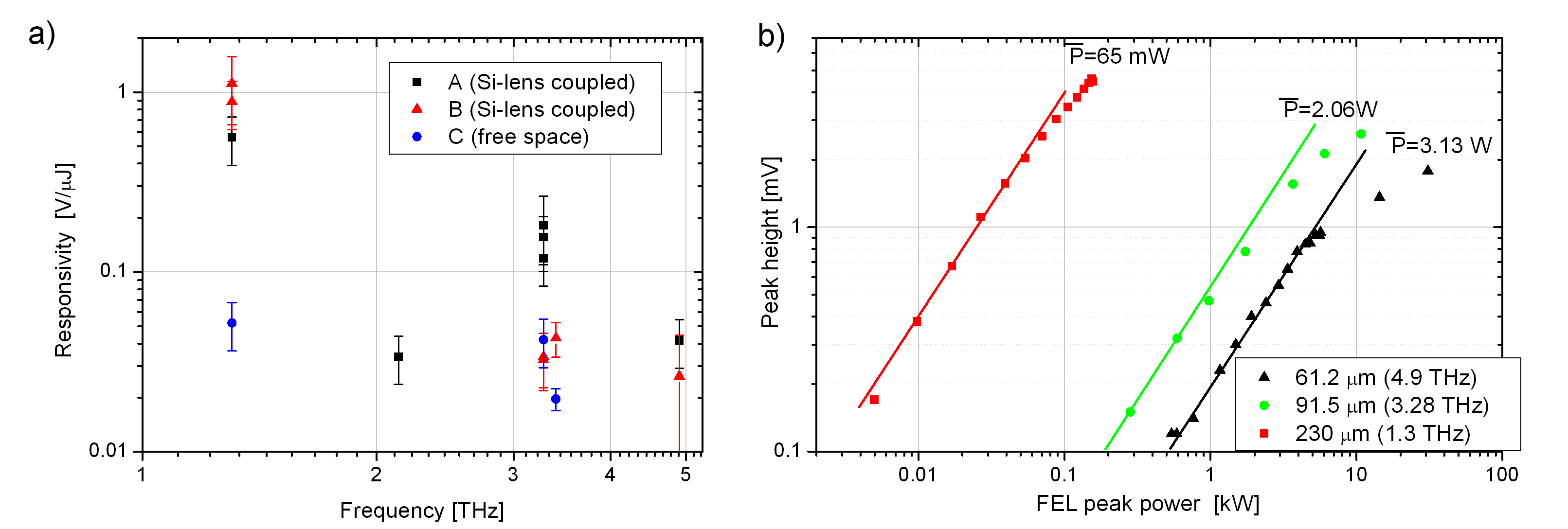}{16 cm}{h}{a) Responsivity of all investigated devices with respect to the pulse energy for a reverse bias of $U_{SG}\approx -0.5\pm 0.1$ V ($U_{G}\approx 0.6$ V). The error bars take the calibration error of the thermal reference detector of an estimated 30\% into account. b) Linearity of sample B. The solid lines indicate a linear THz power-detector signal dependence. The maximum FEL average power is indicated in the graph.}{Fig4}

We now discuss the responsivity and linearity of the devices, which needs consideration of the the post-detection electronics.
Except for the measurement at 1.3 THz, the FEL pulse duration is shorter than the oscilloscope rise time. The oscilloscope cannot resolve the pulse any more, it rather measures the integrated pulse energy than the pulse power, with a fixed HWHM corresponding to the oscilloscope rise time of 17 ps. \Figref{Fig4} a) depicts the pulse energy (E=P$_{THz}^{pk}\tau_{pls}$) responsivity of all investigated devices vs. frequency. Sample C shows a flat response over the investigated frequency range. The responsivity of a rectifying FET is proportional to the radiation resistance of the receiving element.\cite{MeinFETTheo} In ref. \cite{MeinReview} we calculated the radiation resistance of a large area dipole array. It is given by the product of the excited Hertzian dipole of the unit cell of the structure, $R_H(\lambda_{THz}) =80\pi^2\Omega/n_{eff}\times (d/\lambda_{THz})^2$, and the array factor \cite{MeinReview}, resulting in

\begin{equation}
R_A(\lambda_{THz})=R_H(\lambda_{THz})/(1+(0.82\rho k_{THz})^2)=R_H(\lambda_{THz})/(1+(5.15\rho/\lambda_{THz})^2),\label{RA}
\end{equation}

where $\rho$ is the THz spot radius, and $d$ the effective dipole length of an individual mesa. For spot sizes comparable to or larger than the THz wavelength, the radiation resistance for a constant dipole length is frequency-independent, since the wavelength dependences of the Hertzian dipole radiation resistance and the array factor cancel in this limit. Sample C shows such a behaviour.

The other samples show some oscillatory behaviour with lower responsivity at higher frequencies. The reduction of the responsivity with increasing frequency indicates a roll-off due to access resistances \cite{MeinFETTheo}, in agreement with the gate-bias-dependent behaviour of these samples. A similar behaviour has also been found with silicon-based antenna coupled FETs in ref. \cite{Boppel4p3THzBadJourn}, where each antenna-coupled device was optimized for a single operation frequency. Here, a single device can be used for investigating a large frequency range. 
The detectors are well suited for measurements at FELBE and many other FELs: They are most sensitive at the longer wavelengths where the FEL power is small. The smaller responsivity at higher frequencies results in a rather constant dynamic range, since the FEL power increases rapidly with frequency, resulting in similar detected voltages (c.f. \Figref{Fig4} b)). Therefore, the detectors saturate at higher THz power levels.
The oscillatory behaviour of the responsivity may originate from plasmon-resonant effects since the detector is operated in the plasma-wave regime, where the angular THz frequency is much larger than the inverse momentum relaxation time. This criterion is met for THz frequencies above $(2\pi\tau)^{-1}=e/(m^*\mu)\approx 0.5$ THz for high quality GaAs, where $m^*$ is the electron effective mass.  Antenna-less devices provide an ideal platform for investigating plasmonic effects since their radiation resistance shows no pronounced frequency dependence: According to \eqnref{RA}, the radiation resistance is constant for wavelengths shorter than the optical spot diameter and the device size. A further detailed study of the frequency and gate-bias dependence of the LA-FET responsivity may give further insight into plasmon resonances of two dimensional electron gases in the THz range. Such plasmon resonances have yet been used to generate THz radiation from biased field effect transistors \cite{Knap1Emin10}. So far, however, the THz power level was fairly low and the resonances were not very pronounced \cite{KnapFormelEmission}.

\section{Key parameters of the detectors}
The \textit{detection bandwidth} extends down to the lower end of the THz frequency range:
Similar detectors have been tested at 0.48 THz and 0.24 THz at the UCSB-FEL \cite{MeinFETFEL}. The maximum operation frequency of the detector will finally be limited by the Reststrahlenband of GaAs, around 8.3 THz. However, we could not detect any THz power at 7.1 THz any more. We therefore conclude that the LA-FETs are suitable detectors for the major part of the THz frequency range, i.e. from $\sim$ 0.1 THz to at least 5 THz. A single device should be able to cover the whole bandwidth. \Figref{Fig4} b) shows that the detectors are linear up to a detected bias around 2-5 mV, with a larger \textit{linearity range} at lower frequencies. This can be understood by considering the diffraction-limited spot size. At higher frequencies, the THz spot becomes smaller, increasing the intensity in the center of the device, leading to local saturation. The  linearity range is slightly reduced. Taking the measurement bandwidth of the oscilloscope into account, we calculate a \textit{bandwidth-normalized dynamic range} of 65 $\pm$ 3 dB/$\sqrt{Hz}$. This value is in agreement with the results obtained at the UCSB-FEL\cite{MeinFETFEL}. Since the pulse duration was similar to or shorter than the oscilloscope time constant, the \textit{noise equivalent power} (NEP) can only be estimated. Without any amplification and signal conditioning prior to the oscilloscope measurement, the NEP is in the range of 10-60 $\mu$W/$\sqrt{Hz}$ at the longest wavelength of 230 $\mu$m where the pulse duration is comparable to the oscilloscope time constant. This result is limited by the input noise of the oscilloscope. The measurements in ref. \cite{MeinFETFEL} on similar devices revealed an NEP of only 3.1 $\mu$W/$\sqrt{Hz}$. Therefore, the detectors are also \textit{suited for most table-top systems}. Peak power levels of efficient table-top pulsed sources range from W level (average THz power 190 $\mu$W \cite{AreaEmitterWinnerl}) even to the few MW peak power-level \cite{TiltedPulseFront2,DekorsyNeu}. Once calibrated, the LA-FETs will be suitable for power measurements using the lock-in technique. They are much faster than conventional thermal detectors, allowing for fast data acquisition and power monitoring. The detectors presented in ref. \cite{MeinFETFEL} had already been used for continuous-wave power measurements between 0.3 and 0.7 THz.

\section{Conclusion}
The large area field effect transistor rectifiers (LA-FETs) allow for resolving both THz and optical pulses at the tens of ps timescale. These ultrafast detectors are ideally suited for precise timing of optical and THz beams in pump and probe experiments. LA-FETs were sucessfully tested from 0.24 THz to 4.9 THz. We therefore conclude that a single LA-FET may be used to cover most of the THz frequency range. The detectors have been tested up to 30 kW peak power levels without any noticeable thermal limitation. The optical response time ($\sim$ 150 ps) can be further reduced by reducing the size of the mesa features. The measurement of the THz response time was limited by the oscilloscope resolution (< 30 ps). Since the smallest dimension of the devices is 3 $\mu$m, they can be manufactured with simple processing techniques. 
The relatively low NEP and large dynamic range suggest that these detectors can also be used in pulsed and continuous-wave table top systems. 

The author S.P. thanks M. Latzel for deposition of dielectrics. We thank W. Seidel, P. Michel and the FELBE team for their dedicated support. 

\bibliographystyle{osajnl}

\end{document}